\documentclass[10pt,twocolumn,twoside]{ieeeconf}
\IEEEoverridecommandlockouts    
\newtheorem{remark}{Remark}

\newtheorem{lemma}{Lemma}
\newtheorem{prop}{Proposition}

\newtheorem{theorem}{Theorem}
\newtheorem{assumption}{Assumption}

\usepackage{cite}
\usepackage{amsmath}
\usepackage{amstext}
\usepackage{amssymb}
\usepackage{tikz}
\usepackage[mathscr]{euscript}

\usepackage{amsfonts}
\usepackage{enumerate}
\usepackage[final]{pdfpages}
\usepackage{csquotes}

\let\labelindent\relax
\DeclareOldFontCommand{\rm}{\normalfont\rmfamily}{\mathrm}

\usepackage{graphicx}      
\usepackage{subcaption}
\usepackage{enumitem}
\usepackage{mathdots}

\usepackage{mathtools,xparse}
\DeclareMathOperator{\diag}{diag}
\definecolor{ForestGreen}{RGB}{34,139,34}

\def\qed{\hfill $\Box$}

\begin{document}
\author{Sebin~Gracy,
         ~Brian D.O. Anderson,
        ~Mengbin~Ye
	   ~and~C\'esar A.~Uribe
\thanks{Sebin Gracy and C\'esar A.~Uribe are with the Department of Electrical and Computer Engineering, Rice University, Houston, TX, USA (\texttt{sebin.gracy@rice.edu}, \texttt{cesar.uribe@rice.edu}). \\
Brian D.O. Anderson is with the School of Engineering, Australian National University, Canberra, Australia. (\texttt{brian.anderson@anu.edu.au}) Mengbin~Ye is with the Centre for Optimisation and Decision Science, Curtin University, Australia. (\texttt{mengbin.ye@curtin.edu.au}).
}}
\title{\LARGE \bf Competitive Networked Bivirus SIS spread over Hypergraphs}


\maketitle
\begin{abstract}
The paper deals with the spread of two competing viruses over a network of population nodes, accounting for pairwise interactions and higher-order interactions (HOI) within and between the population nodes. We study the competitive networked bivirus susceptible-infected-susceptible (SIS) model on a hypergraph introduced in Cui et al. ~\cite{cui2023general}. We show that the system has, in a generic sense, a finite number of equilibria, and the Jacobian associated with each equilibrium point is nonsingular; the key tool is the Parametric Transversality Theorem of differential topology. Since the system is also monotone, it turns out that the typical behavior of the system is convergence to some equilibrium point. Thereafter, we exhibit a tri-stable domain with three locally exponentially stable equilibria. For different parameter regimes, we establish conditions for the existence of a coexistence equilibrium (both viruses infect separate fractions of each population node). 
\end{abstract}

\section{Introduction}
The study of virus spread has been an active area of research for over two centuries. In particular, diverse scientific  communities, such as physics \cite{van2008virus}, mathematics \cite{castillo1989epidemiological}, computer science \cite{prakash2012winner}, 
automatic control \cite{nowzari2016analysis}
, etc., have significantly aided in furthering our understanding of the complex mechanisms behind the spread of a virus. Fundamental to this effort has been the development of compartmental models where each individual is healthy and susceptible (S), infected with a virus (I), or has recovered from a viral infection (R). Two compartmental models, susceptible-infected-recovered (SIR) and susceptible-infected-susceptible (SIS) have garnered significant attention in several scientific disciplines, particularly in mathematical epidemiology. In contrast to the SIR model, the SIS model allows for the possibility of reinfection and is the focus of the present paper. More specifically, we will deal with networked SIS models, with each node in the network being representative of a large population, and the interconnection among the nodes denotes the possible spreading pathways for the virus.  


The existing literature on modeling virus spread typically relies on the assumption that there is just a single virus present. However, one often encounters scenarios where there are two viruses, say virus~1 and virus~2,  circulating in a meta-population (i.e., a network of population nodes). In such a context,  said viruses could be cooperative, i.e., infection with virus~1 (resp. virus~2) increases (resp. decreases) the likelihood of simultaneous infection with virus~2 (resp. virus~1); see \cite{gracy2022modeling} for more details. Another possibility is for the two viruses to compete; infection with virus~1 (resp.~virus~2) precludes the possibility of simultaneous infection with virus~2 (resp.~virus~1) - this is the focus of the present paper. We stress that the notion of competing viruses is not restricted to just epidemics; it manifests itself in, among others,  product adoption in a marketplace and the spread of opinions in social networks~\cite{nedic2019graph}.

Networked competitive multi-virus SIS models have been analyzed in substantial depth in recent times; see~ \cite{sahneh2014competitive,axel2020TAC,liu2019analysis,ye2021convergence,anderson2023equilibria,santos2015bi,gracy2022siads,castillo1989epidemiological,carlos2}. A major drawback of networked competitive bivirus SIS models studied in the aforementioned papers is that they account only for pairwise interactions between individuals. 
In reality, interactions in social groups often involve more than two individuals - it is not unusual that an individual can \emph{simultaneously} interact with more than one other individual. This motivates the need for higher-order networks such as hypergraphs~\footnote{Simplicial networks (see \cite{hatcher2005algebraic} 
for more details) 
have also been used for studying HOI, see \cite{bick2023higher}.}, i.e., graphs where an edge can connect more than two nodes, which are quite effective in representing higher-order interactions (HOI) \cite{bick2023higher}. Inspired by the approach in \cite{iacopini2019simplicial}, an SIS model on a hypergraph has been proposed and analyzed in~\cite{de2020social}. However, the analytic results therein relied on certain restrictions on the network structure. Overcoming this drawback, a networked SIS model on a hypergraph has been devised and studied in considerable detail in 
\cite{cisneros2021multigroup}. However, the modeling frameworks in \cite{iacopini2019simplicial,de2020social,cisneros2021multigroup} are restrictive in the sense that none of these account for the possibility of more than one virus simultaneously circulating in a given population. Addressing this shortcoming, a competitive networked bivirus SIS model on a hypergraph has been developed and analyzed in~\cite{cui2023general}. The set of equilibria for the model in  \cite{cui2023general} can be broadly classified into three categories: the disease-free equilibrium (both viruses have been eradicated), the boundary equilibria (one virus is dead, and the other is alive); and coexistence equilibria (two viruses infect separate fractions of every population node in the network). Nevertheless, the results in~\cite{cui2023general} have the following limitations: a) some of the findings therein have yet to be rigorously established, and b) the analysis, while improving our understanding of the existence and stability of various equilibria, is not exhaustive. The present paper aims to address the aforementioned gaps.
Our main contributions, therefore, are as follows:
\begin{enumerate}[label=\roman*)]
    \item We show that the networked bivirus SIS system with HOI has, in a generic sense, a finite number of equilibria. Furthermore, for each equilibrium, the associated Jacobian is a nonsingular matrix; see Theorem~\ref{thm:finiteness}. In so doing, since our proof of Theorem~\ref{thm:finiteness} does not, unlike the proof of \cite[Theorem~5.5]{cui2023general}, require the HOI infection rates to be set to zero, we establish the correctness of the claim raised in \cite[Theorem~5.5]{cui2023general}. Building off of Theorem~\ref{thm:finiteness} and leveraging the fact that the system is monotone as identified in \cite[Theorem~5.5]{cui2023general}, we prove that the typical behavior of the bivirus SIS system with HOI is convergence to an equilibrium point; see Theorem~\ref{thm:convergence}.
    \item We identify a parameter regime that not only establishes the existence of three equilibria (a single-virus endemic equilibrium corresponding to virus~1 (resp. virus~2) and the DFE) but also guarantees that all of the said equilibria are locally exponentially stable at the same time; see Proposition~\ref{prop:tristable}. 
    \item We identify a parameter regime, different from the one covered by Proposition~\ref{prop:tristable}, for the existence of a coexistence equilibrium.  We do so under different configurations of the boundary equilibria, viz. both being unstable and both being stable; see Proposition~\ref{prop:1:suff:coexistence} and Theorem~\ref{prop:suff:coexistence:both stable}, respectively. 
\end{enumerate}
Additionally, for the parameter regime covered by Proposition~\ref{prop:tristable}, we establish existence of a coexistence equilibrium; see Proposition~\ref{prop:2:suff:coexistence:less than 1}.

\noindent \textbf{Notation}:
We denote the set of real numbers by $\mathbb{R}$ and the set of nonnegative real numbers by $\mathbb{R}_+$. For any positive integer $n$, we use $[n]$ to denote the set $\{1,2,...,n\}$. 
We use $\textbf{0}$ and $\textbf{1}$ to denote the vectors whose entries all equal $0$ and $1$, respectively, and use $I$ to denote the identity matrix. 
For a vector $x$, we denote the diagonal square matrix with $x$ along the diagonal by $\diag(x)$. For any two vectors $a, b \in \mathbb{R}^n$ we write $a \geq b$ if $a_i \geq b_i$ for all $i \in [n]$, $a>b$ if $a \geq b$ and $a \neq b$, and $a \gg b$ if $a_i > b_i$ for all $i \in [n]$. Likewise, for any two matrices $A, B \in \mathbb{R}^{n \times m}$, we write $A \geq B$ if $A_{ij} \geq B_{ij}$ for all $i \in [n]$, $j \in [m]$, and $A>B$ if $A \geq B$ and $A \neq B$. 
For a square matrix $M$, we use $\sigma(M)$ to denote the spectrum of $M$, $\rho(M)$ to denote the spectral radius of $M$, and $s(M)$ to denote the spectral abscissa of $M$, i.e., $s(M) = \max\{{\rm{Re}}(\lambda) : \lambda \in \sigma(M)\}$. 

A real square matrix $A$ is called Metzler if all its off-diagonal entries are nonnegative.
 A matrix $A$ is said to be an M-matrix if all of its off-diagonal entries are nonpositive, and there exists a constant $c>0$ such that, for some nonnegative $B$ and $c \geq \rho(B)$, $A=cI-B$. All eigenvalues of an M-matrix have nonnegative real parts. Furthermore, if an M-matrix has an eigenvalue at the origin, we say it is singular; if each eigenvalue has strictly positive parts, then we say it is nonsingular. If $A(=[a_{ij}]_{n\times n})$ is a nonnegative   matrix, then $\rho(A)$  decreases monotonically with a decrease in $a_{ij}$ for  any
 $i,j \in [n]$. The matrix $A$ is reducible if, and only if, there is a permutation matrix $P$ such that $P^\top AP$ is block upper triangular; otherwise, $A$ is said to be irreducible. If a nonnegative $A$ is irreducible, and $Ax = y$ for $x > \textbf{0}$, then $y > \textbf{0}$, and $y$ cannot have a zero in every position where $x$ has a zero.

\section{Problem Formulation}\label{sec:prob:formulation}
 \subsection{Model}

Consider a network of $n$ nodes. A node represents a well-mixed\footnote{Well-mixed means that the probability of any two individuals in a node interacting with each other is the same.} population of individuals. We will assume that the size of the population is fixed.  We suppose two viruses, say virus~1 and virus~2, are spreading over such a network. Throughout this paper, we will assume that the two aforementioned viruses are competing. Through pairwise or HOI 
as described in more detail below, an otherwise healthy individual in node $i$ gets infected with virus~1 (resp. virus~2) due to contact with either other individuals in node $i$ who are infected with  virus~1 (resp. virus~2) and/or
 with other individuals in node $j$ (where $j$ is a neighbor of $i$) who are infected with virus~1 (resp. virus~2).
When a single interaction is involved (i.e., between two individuals in node $i$ or between an individual in node $i$ and an individual in node $j$), we say that the infection is caused due to \emph{pairwise interactions}. An individual in node $i$ could also be infected with virus~1 (resp. virus~2) due to \emph{simultaneous} interactions with infected individuals in nodes $j$ and $\ell$, where either a) $j = i,$ and/or $\ell=i$, or b) $j, \ell$ are neighbors of $i$. Such interactions are referred to as \emph{higher-order interactions} (HOI). The notion of competition implies that no individual can be simultaneously infected with virus~1 and virus~2.

 We assume that the pairwise infection (resp. HOI) rate with respect to virus~$k$ is the same for all nodes, denoted by $\beta_1^k$ (resp. $\beta_2^k$) for all $i \in [n]$ and $k \in [2]$ \footnote{Indeed, it is far more natural to have possibly different infection rates for each node; it is standard in the literature on classic SIS bivirus networked systems~\cite{liu2019analysis,sahneh2014competitive,axel2020TAC,ye2021convergence,anderson2023equilibria,pare2021multi,santos2015bi}. As evident below, we do not impose constraints on the values of the nonnegative matrices capturing the interactions, and hence the analysis does not differ materially. We choose this particular notation to remain consistent with earlier literature on epidemic models with HOI~\cite{cisneros2021multigroup}.
 }. An individual infected with virus~$k$ recovers from said infection at a healing rate $\delta_i^k$ and immediately becomes susceptible to virus~1 or by virus~2. All individuals within a node have the same healing rate with respect to virus~$k$; individuals in different nodes possibly have different healing rates. We say that node $i$ is healthy if all individuals in node $i$ are healthy; otherwise, we say it is infected. Within the same node, it is possible for there to simultaneously exist a fraction of individuals that are infected with virus~$1$ and for a different fraction that is infected with virus~$2$.

As mentioned previously, diseases could spread due to pairwise interactions and HOI. In case of the former, if an individual in node $j$ can infect an individual in node $i$ with virus~$k$, then, with $a_{ij}^k (\geq 0)$ denoting the strength of interactions between an individual in node $j$ and an individual in node $i$ with respect to spread of virus~$k$,   we have that $a_{ij}^k>0$; otherwise $a_{ij}^k=0$. For the case of HOI, if an individual in node $i$ gets infected with virus~$k$ due to simultaneous interactions with individuals in nodes $j$  and $\ell$, then, with $b_{ij\ell}^k$ denoting the strength of interaction that nodes $j$ and $\ell$ together have on node $i$ with respect to the spread of virus~$k$, we have that $b_{ij\ell}^k>0$; else, $b_{ij\ell}^k=0$. Let $x_i^k(t)$ denote the fraction of individuals infected with virus~$k$ in agent~$i$ at time instant $t$. The evolution of this fraction can, therefore, be represented by the following scalar differential equation \cite[Section~5]{cui2023general}, where, for $i=1,2,\hdots,n$, we have
\begin{align}\label{eq:ct:hypergraph}
\dot{x}_i^1=& -\delta_i^1x_i^1+\beta_1^1(1-x_i^1-x_i^2)\textstyle\sum_{j=1}^{n}a_{ij}^1x_j^1 + \nonumber \\ &~~
\beta_2^1(1-x_i^1-x_i^2)\textstyle\sum_{j,\ell=1}^{n}b_{ij\ell}^1x_{j}^1x_{\ell}^1 \nonumber\\
\dot{x}_i^2=&-\delta_i^2x_i^2+\beta_1^1(1-x_i^1-x_i^2)\textstyle\sum_{j=1}^{n}a_{ij}^2x_j^2 +\nonumber \\ &~~
\beta_2^2(1-x_i^1-x_i^2)\textstyle\sum_{j,\ell=1}^{n}b_{ij\ell}^2x_{j}^2x_{\ell}^2 
\end{align}
\normalsize
Define $D^1 =\diag(\delta_i^1)$, where $i\in [n]$, and define $D^2$ analogously. Define $X^1=\diag(x_i^1)$,  where $i\in[n]$, and define $X^2$ analogously. Let $A^1=[a_{ij}^1]_{n \times n}$, and $A^2=[a_{ij}^2]_{n \times n}$. Let $B_i^k=[b_{ij\ell}^k]_{n \times n}$, for each $i \in [n]$ and $k \in [2]$. Let $x^k =\small \begin{bmatrix}x_1^k &&x_2^k&&\hdots&& x_n^k \end{bmatrix}^\top$ for $k=1,2$. 

Therefore, in vector form, equation~\eqref{eq:ct:hypergraph} can be written as:
\footnotesize
\begin{align}\label{eq:bivirus:hypergraph}
\dot{x}^1=&-D^1x^1 + \beta_1^1(I-X^1-X^2)A^1x^1 + \nonumber \\ 
&~~~~\beta_2^1(I-X^1-X^2)
((x^1)^\top B_1^1x^1, (x^1)^\top B_2^1x^1, \hdots, (x^1)^\top B_n^1x^1)^\top \nonumber \\
\dot{x}^2=&-D^2x^2 + \beta_1^2(I-X^1-X^2)A^2x^2 + \nonumber \\
&~~~~\beta_2^2(I-X^1-X^2)
((x^2)^\top B_1^2x^2, (x^2)^\top B_2^2x^2, \hdots, (x^2)^\top B_n^2x^2)^\top
\end{align}
 \normalsize 

Throughout this document, we will drop the superscript~$k$ while considering the single-virus case.  

We note that system~\eqref{eq:bivirus:hypergraph} is a special case of \cite[system~5.5]{cui2023general} in the following sense: System~\eqref{eq:bivirus:hypergraph} only accounts for a) the case where, for $k=1,2$, $\beta_i^k$ and $\beta_2^k$ is identical for every node $i$, $i=1,2,\hdots, n$, and b) the case where virus~1 (resp.~virus~2) spread only due to contact with the infected individuals. In contrast, the model in \cite{cui2023general} (see \cite[system~5.5]{cui2023general}) allows for the possibility of $\beta_i^k$ and $\beta_2^k$ being not necessarily the same for every node. Furthermore, it also allows for the possibility of the viruses to spread through additional mediums such as a water distribution network, a public transit network, etc.

\begin{remark}\label{rem:special:case}
    Note that setting $\beta_2^k=0$ for $k=1,2$ results in system~\eqref{eq:bivirus:hypergraph} coinciding with the classic networked bivirus SIS model studied in, among others, \cite{liu2019analysis,ye2021convergence,axel2020TAC,sahneh2014competitive,santos2015bi,anderson2023equilibria}. Setting $x^1(0)=\textbf{0}$ (resp.~$x^2(0)=\textbf{0}$) results in system~\eqref{eq:bivirus:hypergraph} coinciding with the model used for studying the spread of a single virus over hypergraphs in~\cite{cisneros2021multigroup}.
\end{remark}

\noindent The model in system~\eqref{eq:bivirus:hypergraph} has three kinds of equilibria, viz. healthy state or disease-free equilibrium (DFE), $(\textbf{0}, \textbf{0})$; single-virus endemic equilibria corresponding to virus~$k$, of the form $(\bar{x}^k, \textbf{0})$, where $\textbf{0}\ll \bar{x}^k\ll \textbf{1}$ for $k=1,2$; and coexisting equilibria, $(\bar{x}^1, \bar{x}^2)$, where, as we will show in Lemma~\ref{lem:equi_non-zero_nonone:1}, $\textbf{0}\ll \bar{x}^1, \bar{x}^2 \ll \textbf{1}$, and, furthermore, $\bar{x}^1+ \bar{x}^2 \ll \textbf{1}$. 
 It is unknown whether the single-virus endemic equilibria corresponding to virus~$k$  are unique, in contrast to the classic bivirus SIS network model without HOI.

\noindent The Jacobian of system~\eqref{eq:bivirus:hypergraph} evaluated at an arbitrary point, $(x^1, x^2)$, in the state space is as given in~\eqref{eq:jacob}.
\begin{equation}\label{eq:jacob}
J(x^1, x^2) = 
\begin{bmatrix}
J_{11} && J_{22}\\
J_{21} && J_{22}
\end{bmatrix},
\end{equation}
where \footnotesize
\begin{align}
J_{11} =&-D^1+\beta_1^1(I-X^1-X^2)A^1 - \diag(\beta_1^1A^1x^1)+  \nonumber \\ &~~~~\beta_2^1(I-X^1-X^2)O_1(x^1) -\beta_2^1O_2(x^1)\label{eq:J11}\\
J_{12} =& -\diag(\beta_1^1A^1x^1) - \beta_2^1\diag((x^1)^\top B_i^1x^1)_{i=1,2,\hdots, n}\label{eq:J12}\\
J_{21} = &-\diag(\beta_1^2A^2x^2) - \beta_2^2\diag((x^2)^\top B_i^2x^2)_{i=1,2,\hdots, n}\label{eq:J21}\\
J_{22} = &-D^2+\beta_1^2(I-X^1-X^2)A^2 - \diag(\beta_1^2A^2x^2)+  \nonumber \\ &~~~~\beta_2^2(I-X^1-X^2)O_3(x^2) -\beta_2^2O_4(x^2)\label{eq:J22}
\end{align}

\normalsize 
\noindent 
The terms $O_1(x^1)$, $O_2(x^1)$, $O_3(x^2)$ and $O_4(x^2)$ are as given in~\eqref{eq:O1},~\eqref{eq:O2},~\eqref{eq:O3} and~\eqref{eq:O4}, respectively. 
\begin{figure*}
\begin{align}
O_1(x^1) = & \begin{bmatrix} (B_1^1+(B_1^1)^\top)x^1 && (B_2^1+(B_2^1)^\top)x^1 && \hdots && (B_n^1+(B_n^1)^\top)x^1\end{bmatrix}\label{eq:O1} \\
O_2(x^1) = & \diag((x^1)^\top B_i^1 x^1)_{i=1,2,\hdots, n} \label{eq:O2} \\
O_3(x^2) =  &\begin{bmatrix} (B_1^2+(B_1^2)^\top)x^2 && (B_2^2+(B_2^2)^\top)x^2 && \hdots && (B_n^2+(B_n^2)^\top)x^2\end{bmatrix} \label{eq:O3} \\
O_4(x^2) =& \diag((x^2)^\top B_i^2 x^2)_{i=1,2,\hdots, n} \label{eq:O4}
\end{align}
\end{figure*}
\normalsize 
\noindent We will need the following assumptions to ensure the model is well-defined.
\begin{assumption}\label{assum:1}
    The matrix $D^k$, for $k=1,2$, is a positive diagonal matrix. The matrix $A^k$, for $k=1,2$, is nonnegative. The matrix $B_i^k$ is nonnegative for all $i \in [n]$ and $k \in [2]$.
\end{assumption}

\begin{assumption}\label{assum:2}
    The matrix $A^k$, for $k=1,2$, is  irreducible.
\end{assumption}

\noindent We define the set $\mathcal D$ as follows: \begin{equation}\label{eq:D}\mathcal D: = \{(x^1, x^2)\mid x^k \geq \textbf{0}, k=1,2, \textstyle\sum_{k=1}^2x^k\leq \mathbf{1}\}.\end{equation}

It is known that the set $\mathcal D$ is positively invariant, and that the DFE is always an equilibrium for system~\eqref{eq:bivirus:hypergraph}; see \cite[Lemma~5.1]{cui2023general}. The fact that $\mathcal D$ is positively invariant guarantees that the state values $x^k_i, k\in[2], i\in[n]$, always stay in the $[0,1]$ interval. Since the states represent fractions of an infected population node, if the states were to take values outside the $[0,1]$ interval, then those would not correspond to physical reality.

\subsection{Problem Statements}\label{sec:prob:statement}
With respect to system~\eqref{eq:bivirus:hypergraph}, we aim to answer the following questions in this paper conclusively:
\begin{enumerate}[label=\roman*)]
\item \label{qs0} What is the typical behavior the trajectories exhibit as time goes to infinity?
\item \label{qs1} Can we identify a parameter regime such that multiple equilibria are simultaneously stable?

\item  \label{qs2} 
Can we identify sufficient conditions for the existence of a coexistence equilibrium? 
Furthermore, can we establish the stability properties of such an equilibrium based on knowledge of the stability properties of the boundary equilibria?

\end{enumerate}
\subsection{Preliminary Lemmas and analysis of healthy state}
In this subsection, we will establish certain preliminary results on the nature of equilibria of system~\eqref{eq:ct:hypergraph}, and recall some of the results on irreducible matrices - all of these will aid in the development of the main results of the paper.

\begin{lemma} \label{lem:equi_non-zero_nonone:1}
    Consider system~\eqref{eq:bivirus:hypergraph} under Assumptions~\ref{assum:1} and~\ref{assum:2}. 
 If $\bar{x} = (\bar{x}^1,  \bar{x}^2) \in \mathcal{D}$ is an equilibrium of~\eqref{eq:bivirus:hypergraph}, then, for each $k \in [2]$, either $\bar{x}^k = \textbf{0}$, or $\textbf{0} \ll \bar{x}^k \ll \textbf{1}$. Moreover, 
$\textstyle \sum_{k=1}^2 \bar{x}^k \ll \textbf{1}$.
\end{lemma} 
The proof is inspired from \cite[Lemma~3.1]{ye2021convergence}.\\
\noindent \textit{Proof:} 
It is clear that $(\textbf{0}, \textbf{0})$ is an equilibrium of~\eqref{eq:bivirus:hypergraph}. Therefore, in the rest of the proof, we will show that any non-zero equilibrium $\bar{x} = (\bar{x}^1,  \bar{x}^2)$ of~\eqref{eq:bivirus:hypergraph} must satisfy, for each $k \in [2]$, $\textbf{0} \ll \bar{x}^k \ll \textbf{1}$ and $\textstyle \sum_{k=1}^2 \bar{x}^k \ll \textbf{1}$. 
\noindent We start off by showing that $\bar{x}^1 + \bar{x}^2 \ll \textbf{1}$.  For any $i \in [n]$, observe that the following is satisfied:
\begin{align}\label{eq:ct:hypergraph:equilibrium:version:3}
\dot{\bar{x}}_i^1+\dot{\bar{x}}_i^2=&-\delta_i^1\bar{x}^1_i -\delta_i^2\bar{x}^2_i + \beta_1^1(1-\bar{x}_i^1-\bar{x}_i^2)\textstyle\sum_{j=1}^{n}a_{ij}^1\bar{x}_j^1 \nonumber \\ &~
+\beta_2^1(1-\bar{x}_i^1-\bar{x}^2_i)\textstyle\sum_{j,\ell=1}^{n}b_{ij\ell}^1\bar{x}_{j}^1\bar{x}_{\ell}^1 \nonumber\\ &~+ \beta_1^2(1-\bar{x}_i^1-\bar{x}_i^2)\textstyle\sum_{j=1}^{n}a_{ij}^2\bar{x}_j^2 \nonumber \\&~
+\beta_2^2(1-\bar{x}_i^1-\bar{x}^2_i)\textstyle\sum_{j,\ell=1}^{n}b_{ij\ell}^2\bar{x}_{j}^2\bar{x}_{\ell}^2
\end{align}
Suppose that, for some $i \in [n]$, $\bar{x}_i^1+\bar{x}^2_i =1$. Therefore, since, by Assumption~\ref{assum:1}, $\delta_i^k>0$ for $k=1,2$, and since $\bar{x}_i^k \in \mathcal D$, from~\eqref{eq:ct:hypergraph:equilibrium:version:3}, it is clear that $\dot{\bar{x}}_i^1+\dot{\bar{x}}_i^2<0$. However, since by assumption $\bar{x}=(\bar{x}^1, \bar{x}^2)$ is an equilibrium, it must be that $\dot{\bar{x}}_i^1+\dot{\bar{x}}_i^2=0$, which is a contradiction. Therefore, for all $i \in [n]$, $\bar{x}_i^1+\bar{x}^2_i <1$, which implies that $\textstyle \sum_{k=1}^2 x^k \ll \textbf{1}$; thus guaranteeing that $\bar{x}^k \ll \textbf{1}$ for $k=1,2$.\\
We are left to show that $\bar{x}^k \gg \textbf{0}$ for $k=1,2$. To this end, suppose that $\bar{x}^1>\textbf{0}$ is an equilibrium point for which there exists at least one (but possibly more) $i \in [n]$ such that $\bar{x}_i^1=0$. Note that the equilibrium version of the first line of equation~\eqref{eq:bivirus:hypergraph} yields the following:
\footnotesize
\begin{align}\label{eq:ct:hypergraph:equilibrium:version}
\dot{\bar{x}}^1=&-D^1\bar{x}^1 + \beta_1^1(I-\bar{X}^1-\bar{X}^2)A^1\bar{x}^1 + \nonumber \\ 
&~~~~\beta_2^1(I-\bar{X}^1-\bar{X}^2)
((\bar{x}^1)^\top B_1^1\bar{x}^1, (\bar{x}^1)^\top B_2^1\bar{x}^1, \hdots, (\bar{x}^1)^\top B_n^1\bar{x}^1)^\top 
\end{align}
\normalsize 
By noting that $\bar{x}^1$ is an equilibrium point, and by a suitable rearrangement of terms, we obtain:
\begin{align}\label{eq:ct:hypergraph:equilibrium:version:2}
\bar{x}^1=& S \bar{x}^1,
\end{align}
\normalsize where
\footnotesize 
\begin{align}
S= &(D^1)^{-1}\beta_1^1(I-\bar{X}^1-\bar{X}^2)A^1 +  \nonumber \\ 
~~~~&(D^1)^{-1}\beta_2^1(I-\bar{X}^1-\bar{X}^2)
((\bar{x}^1)^\top B_1^1, 
\hdots, (\bar{x}^1)^\top B_n^1.
 \end{align}
\normalsize 
By Assumptions~\ref{assum:1} and~\ref{assum:2}, it is clear that the matrix $S$ is nonnegative and irreducible.
Since, by assumption, $\bar{x}^1 > \textbf{0}$, from~\eqref{eq:ct:hypergraph:equilibrium:version:2} and coupled with a property of a nonnegative matrix that is irreducible 
 we have the following: i) $S\bar{x}^1 > \textbf{0}$, and ii) there is at least one $i \in [n]$ such that $\bar{x}^1_i=0$ but $(S\bar{x}^1)_i>0$. Note that ii) 
contradicts~\eqref{eq:ct:hypergraph:equilibrium:version:2}. Therefore, if $\bar{x}^1 > \textbf{0}$ is an equilibrium point, then it must be that $\bar{x}^1 \gg \textbf{0}$. By an analogous argument, it can be shown that $\bar{x}^2 \gg \textbf{0}$, thus completing the proof.
~\qed

 \begin{lemma}\cite[Proposition 1]{liu2019analysis}] \label{lem:eigspec}
Suppose that $\Lambda$ is a negative diagonal matrix and $N$ is an irreducible nonnegative matrix. 
Let $M$ be the irreducible Metzler matrix $M = \Lambda+N$. 
Then, $s(M) < 0$ if and only if $\rho(-\Lambda^{-1} N) < 1, s(M)=0$ if and only if $\rho(-\Lambda^{-1} N) = 1$, and $s(M)>0$ if and only if, $\rho(-\Lambda^{-1} N) > 1$.
\end{lemma}

\begin{lemma}\cite[Proposition~2]{rantzer}\label{lem:ratzner}
Let $A \in \mathbb{R}^{n\times n}$ be Metzler. Then, $A$ is Hurwitz if, and only if, there exists an $x \in \mathbb{R}^n$ such that $ x \gg \textbf{0}$ and $Ax \ll 0$.
\end{lemma}

\begin{lemma} \label{lem:perron_frob}
\cite[Chapter 8.3]{meyer2000matrix} \cite[Theorem~2.7]{varga1999matrix} 
Suppose that $N$ is an irreducible nonnegative matrix. Then,
\begin{enumerate}[label=(\roman*)]
    \item $r = \rho(N)$ is a simple eigenvalue of $N$. \label{item:perfrob_simpleeig}
    \item There is an eigenvector $\zeta \gg \textbf{0}$ corresponding to the eigenvalue $r$. \label{item:perfrob_pos_exists}
    \item $x > \textbf{0}$ is an eigenvector only if $Nx = rx$ and $x \gg \textbf{0}$. 
    \label{item:perfrob_pos_necess}
    \item If $A$ is a nonnegative matrix such that $A < N$, then $\rho(A) < \rho(N)$. \label{item:perfrob_matrix_ineq}~$\blacksquare$
\end{enumerate}
\end{lemma}

It can be seen that  $(\textbf{0}, \textbf{0})$ is an equilibrium of~\eqref{eq:bivirus:hypergraph}, and is referred to as the disease-free equilibrium (DFE). We recall a sufficient condition for convergence to the DFE. 
\begin{lemma}\cite[Theorem~5.2, statement~1]{cui2023general} \label{lem:local:stab:DFE}
   Consider system~\eqref{eq:bivirus:hypergraph} under Assumptions~\ref{assum:1} and~\ref{assum:2}. If, for $k=1,2$, $\rho(\beta_1^k(D^k)^{-1}A^k)<1$, then the DFE is locally stable.
\end{lemma}
\noindent Note that the guarantees provided by Lemma~\ref{lem:local:stab:DFE} are only local. It turns out that the DFE, under appropriate conditions, is endowed with stronger stability guarantees. We define for $k=1,2$ the following matrices: 
$$R^k: = \tiny \begin{bmatrix} \textbf{1}^\top B_1^k \\ \textbf{1}^\top B_2^k \\ \vdots \\ \textbf{1}^\top B_n^k \end{bmatrix}.$$
\normalsize With the matrices $R^k$, $k=1,2$, in hand,  we can  recall the following result. 
\begin{lemma}\cite[Theorem~5.2, statement~2]{cui2023general} \label{lem:global:stab:DFE}
   Consider system~\eqref{eq:bivirus:hypergraph} under Assumptions~\ref{assum:1} and~\ref{assum:2}. If, for $k=1,2$, $\rho(\beta_1^k(D^k)^{-1}A^k + \beta_2^k(D^k)^{-1}R^k)<1$, then the DFE is globally exponentially stable.
\end{lemma}

\section{Monotone dynamical systems and competitive bivirus networked SIS models with HOI}\label{sec:monotone}
Monotone dynamical systems (MDS) are a class of 
systems that has found resonance in mathematical epidemiology; one of the major reasons for this is the fact that MDS, assuming that they generically have a finite number of equilibria, converge to a (stable) equilibrium point for almost all initial conditions. Here, the term \enquote{almost all} means: for all but a set of parameter values that has measure zero. An algebraic or semi-algebraic set defines this set of exceptional values. 
It is known that  under Assumptions~\ref{assum:1} and~\ref{assum:2}, system~\eqref{eq:bivirus:hypergraph} is monotone; see \cite[Theorem~5.5]{cui2023general}.  That is,  suppose that $(x_A^1(0), x_A^2(0))$ and $(x_B^1(0), x_B^2(0))$ are two initial conditions in $\textrm{int}(\mathcal D)$ satisfying i) $x_A^1(0)>x_B^1(0)$ and ii) $x_A^2(0)<x_B^2(0)$. Since system~\eqref{eq:bivirus:hypergraph} is monotone, it follows that, for all $t \in \mathbb{R}_{\geq 0}$, i) $x_A^1(t)\gg x_B^1(t)$ and ii) $x_A^2(t)\ll x_B^2(t)$. Note that \cite[Theorem~5.5]{cui2023general} also claims that system~\eqref{eq:bivirus:hypergraph}, in a generic sense, has a finite number of equilibria. However, the proof for said assertion is the proof for the case where $\beta_2^k=0$, for $k=1,2$, in \cite[Theorem~3.6]{ye2021convergence}, which is only a  nongeneric case. Since the proof for finiteness of equilibria in~\cite[Theorem~5.5]{cui2023general} is not complete, it leaves open the issue of generic convergence to an equilibrium point.
To remedy this, we provide a different proof for generic finiteness of equilibria that does not rely on $\beta_2^k=0$ for $k=1,2$.  


Given that nonlinear systems can have complex equilibria patterns, including a continuum of equilibria for the classic bivirus network model, 
we establish that for generic parameter matrices, system~\eqref{eq:ct:hypergraph} has a finite number of equilibria. 
We use arguments very much like those in \cite{anderson2023equilibria}. Essentially because the healthy equilibrium and the single-virus boundary equilibria can be conveniently studied using single-virus techniques, it is easily established that there are no continua of equilibria confined to any boundary, i.e. any continuum of equilibria necessarily includes a continuum of coexistence equilibria. Therefore, we focus on showing that such equilibria cannot exist for generic parameter values. The tool is the Parametric Transversality Theorem, see \cite[see p. 145]{lee2013introduction} and \cite[see p.68]{guillemin2010differential}. The main result is as follows:

\begin{theorem}\label{thm:finiteness}
Consider the model of \eqref{eq:bivirus:hypergraph}, under Assumptions \ref{assum:1} and \ref{assum:2}. With any fixed matrices $A^k$ and nonnegative $B^k_i$, and the exclusion of a set of values for the entries of $D^1,D^2$ of measure zero, the number of coexistence equilibrium points is finite, and the associated vector field zero is nondegenerate, i.e. the associated Jacobian is nonsingular. Similarly, with any fixed $D^1,D^2$ and $B^k_i$, and the exclusion of a set of values for the entries of $A^1,A^2$ of measure zero, the same properties of equilibrium points hold.
\end{theorem}

\noindent \textit{Proof:} See Appendix.~\qed

Theorem~\ref{thm:finiteness}, coupled with the fact that system~\eqref{eq:bivirus:hypergraph} is monotone, allows us to leverage Hirsch's generic convergence theorem \cite{hirsch1985systems} to draw conclusions on the limiting behavior of system~\eqref{eq:bivirus:hypergraph} 
 outside of the specific conditions identified in Lemma~\ref{lem:local:stab:DFE}. 
 We have the following result.
\begin{theorem}\label{thm:convergence}
Consider system~\eqref{eq:bivirus:hypergraph}
 under Assumptions~\ref{assum:1} and ~\ref{assum:2}. 
For all initial
conditions $(x^1(0), x^2(0)) \in \mathcal D$   except possibly for a set of measure zero, the system~\eqref{eq:bivirus:hypergraph} will converge to an equilibrium. If the system does not converge to an equilibrium,
it is on a nonattractive limit cycle.
\end{theorem}
\medskip In words, Theorem~\ref{thm:convergence} establishes that the typical behavior of system~\eqref{eq:bivirus:hypergraph} is convergence to \emph{some} equilibrium; this could be healthy, or (one of the possibly many) single-virus boundary equilibria, or a coexistence equilibrium. It further says that limit cycles, if any, are nonattractive. No more complicated behavior is allowed; chaos can be ruled out, see \cite{sontag2007monotone}. Thus, Theorem~\ref{thm:convergence} answers question~\ref{qs0} raised in Section~\ref{sec:prob:formulation}.\\
Theorem~\ref{thm:convergence} strengthens the result in \cite[Theorem~3.6]{ye2021convergence} by extending the generic convergence behavior to bi-virus SIS models 
that also account for HOI. Furthermore, it establishes the correctness of a similar claim raised in \cite[Theorem~5.5]{cui2023general}.

\section{Existence and local stability of boundary 
equilibria}\label{sec:boundary:equilibria:stability}


In this section, we identify a parameter regime that permits three equilibria of the bivirus system~\eqref{eq:bivirus:hypergraph} 
to be simultaneously locally exponentially stable. Subsequently, for a parameter regime different from the one mentioned above, we identify a condition for the existence and instability of a boundary equilibrium. Finally, when there is only one virus, we identify a condition for the existence and local exponential stability of an endemic equilibrium.
\begin{prop}\label{prop:tristable}
Consider system~\eqref{eq:bivirus:hypergraph} under Assumptions~\ref{assum:1} and~\ref{assum:2}, and $B_i^k\geq 0$ for all $i \in [n]$ and $k \in [2]$. Define, for $k=1,2$, $\mathbf{1}_{B^k} \in \{0,1\}^n$ by $(\mathbf{1}_{B^k})_i=1$ if $B_i^k \neq \mathbf{0}$; otherwise $(\mathbf{1}_{B^k})_i=0$. Suppose that the following conditions are fulfilled for $k=1,2$:
    \begin{enumerate}[label=\alph*)]
    \item\label{q1} $\rho(\beta_1^k(D^k)^{-1}A^k)<1$, and
    \item\label{q2} $\min\limits_{i. s.t. B_i^k \neq \mathbf{0}}\big(\frac{\beta_1^k}{\delta_i^k} (A^k\mathbf{1}_{B^k})_i + \frac{\beta_2^k}{2\delta_i^k}\mathbf{1}_{B^k}^\top B_i \mathbf{1}_{B^k} \big)>2$.
    \end{enumerate}
    Then, the following statements are true:
    \begin{enumerate}[label=\roman*)]
    \item\label{r1} The DFE is locally exponentially stable.
    \item\label{r2} there exist equilibria $\bar{x}^k \gg \textbf{0}$ such that $\bar{x}_i^k \geq \frac{1}{2}$ for $k=1,2$, for any $i$ such that $B_i^k \neq \mathbf{0}$.
    \item\label{r3} Any such equilibrium point $(\bar{x}^1, \textbf{0})$ is locally exponentially stable; and 
    \item\label{r4} Any such equilibrium point $(\textbf{0}, \bar{x}^2)$ is locally exponentially stable.
    \end{enumerate}
\end{prop}
\noindent The proof is inspired from \cite[Theorem~5.1, statements~iv) and~v)]{cisneros2021multigroup}.\\
\noindent \textit{Proof of statement~\ref{r1}:} Note that the Jacobian evaluated at the DFE is as follows:
\begin{align}
J(\textbf{0}, \textbf{0}) = 
\scriptsize
\begin{bmatrix}
-D^1+\beta_1^1A^1 & \textbf{0}    \\
\textbf{0}  & -D^2+\beta_1^2A^2 \end{bmatrix}.\normalsize\nonumber
\end{align}
By assumption, $\beta_1^k\rho(D^k)^{-1}A^k)<1$, for $k=1,2$. Therefore, from Lemma~\ref{lem:eigspec}, it must be that $s(-D^k+\beta_1^kA^k)<0$ for $k=1,2$, which, since $J(\textbf{0}, \textbf{0})$ is a block diagonal matrix, and since the matrices $-D^1+\beta_1^1A^1$ and $-D^2+\beta_1^2A^2$ are the only blocks along the main diagonal, 
implies that  $s(J(\textbf{0}, \textbf{0}))<0$. Local 
exponential 
stability of the DFE, then, follows from \cite[Theorem 4.15 and Corollary~4.3]{khalil2002nonlinear}.\\

\noindent \textit{Proof of statement~\ref{r2}:} See \cite[Theorem~5.1, statement~iv)]{cisneros2021multigroup}.\\

\noindent \textit{Proof of statement~\ref{r3}:} Consider the equilibrium point $(\bar{x}^1, \textbf{0})$, and observe that the Jacobian evaluated at this equilibrium is as follows:
\begin{equation}\label{jac:boundary}
J(\bar{x}^1, \textbf{0}) =\begin{bmatrix} \bar{J}_{11} && \bar{J}_{12} \\ \textbf{0} && \bar{J}_{22} \end{bmatrix},
\end{equation}
where 
\begin{align}
\bar{J}_{11} &= -D^1+\beta_1^1(I-\bar{X}^1)A^1 - \diag(\beta_1^1A^1\bar{x}^1) + \nonumber \\ &~~~~~~\beta_2^1(I-\bar{X}^1)O_1(\bar{x}^1) -\beta_2^1O_2(\bar{x}^1) \nonumber \\
\bar{J}_{12} &= -\diag(\beta_1^1A^1\bar{x}^1) -\beta_2^1\diag((\bar{x}^1)^\top B_i^1\bar{x}^1)_{i=1,\hdots, n}\nonumber \\
\bar{J}_{22} &=  -D^2+\beta_1^2(I-\bar{X}^1)A^2. \nonumber 
\end{align}


\noindent The terms $O_1(\bar{x}^1)$ and $O_2(\bar{x}^1)$ are as defined in~\eqref{eq:O1} and~\eqref{eq:O2}.
 We will establish the exponential stability of the 11 and 22 blocks (i.e., $\bar{J}_{11}$ and $\bar{J}_{22}$) separately. 
Observe that 
\footnotesize 
\begin{align}
\bar{J}_{11}=&-D^1+\beta_1^1(I-\bar{X}^1)A^1 - \diag(\beta_1^1A^1\bar{x}^1) \hspace{2mm} + \nonumber \\ &~~~ \beta_2^1(I-\bar{X}^1)O_1(\bar{x}^1) - \beta_2^1\begin{bmatrix} \tiny (\bar{x}^1)^\top B_1^1 \bar{x}^1& &  \\& \ddots& \\ & &  (\bar{x}^1)^\top B_n^1 \bar{x}^1\end{bmatrix}. \nonumber 
\end{align}
\normalsize 
Define summands 
\footnotesize 
\begin{align}
    Q_1:=&-D^1+\beta_1^1(I-\bar{X}^1)A^1+\beta_2^1(I-\bar{X}^1)\begin{bmatrix} \tiny (\bar{x}^1)^\top B_1^1\\ \vdots \\(\bar{x}^1)^\top B_n^1\\ \end{bmatrix}, \text{ and }\nonumber \\
Q_2:=&\beta_2^1(I-\bar{X}^1)\begin{bmatrix} \tiny (\bar{x}^1)^\top (B_1^1)^\top\\ \vdots \\(\bar{x}^1)^\top (B_n^1)^\top \\ \end{bmatrix} - \nonumber \\ &~~~ \diag(\beta_1^1A^1\bar{x}^1) - \beta_2^1\begin{bmatrix} \tiny (\bar{x}^1)^\top B_1^1 \bar{x}^1& &  \\& \ddots& \\ & &  (\bar{x}^1)^\top B_n^1 \bar{x}^1\end{bmatrix}. \nonumber 
\end{align}
\normalsize 
It is immediate that $\bar{J_{11}}=Q_1+Q_2$, which implies that $\bar{J}_{11}\bar{x}^1=Q_1\bar{x}^1+Q_2\bar{x}^1$. Since $\bar{x}^1$ is a single-virus endemic equilibrium corresponding to virus~1, by taking recourse to the equilibrium version of the first line of equation~\eqref{eq:bivirus:hypergraph}, it is clear that $Q_1\bar{x}^1 =\textbf{0}$. Hence, $\bar{J}_{11}\bar{x}^1=Q_2\bar{x}^1$.\\

\noindent Note that 
\footnotesize 
\begin{align}
Q_2\bar{x}^1 &= \beta_2^1(I-\bar{X}^1)\begin{bmatrix} \tiny (\bar{x}^1)^\top (B_1^1)^\top \bar{x}^1\\ \vdots \\(\bar{x}^1)^\top (B_n^1)^\top \bar{x}^1\\ \end{bmatrix} - \nonumber \\ &~~~\diag(\beta_1^1A^1\bar{x}^1)\bar{x}^1 - \beta_2^1\begin{bmatrix} \tiny (\bar{x}^1)^\top B_1^1 \bar{x}^1& &  \\& \ddots& \\ & &  (\bar{x}^1)^\top B_n^1 \bar{x}^1\end{bmatrix}\bar{x}^1. \label{eq:Q2:key}
\end{align}
\normalsize 
Denote by $(Q_2\bar{x}^1)_i$ the $i^{th}$ entry of the vector $Q_2\bar{x}^1$. Therefore, in view of~\eqref{eq:Q2:key}, we have the following:
\begin{equation}\label{eq:Q2:key:1}
    (Q_2\bar{x}^1)_i = -\beta_1^1\Big(\textstyle\sum_{j=1}^{n}a_{ij}^1\bar{x}^1_j\Big)\bar{x}_i^1 +\beta_2^1 (1-2\bar{x}_i^1)((\bar{x}^1)^\top B_i^1 \bar{x}^1)   
\end{equation}
We consider its sign under two circumstances. Suppose first that $B_i^1 =\mathbf{0}$. Then, in view of~\eqref{eq:Q2:key:1}, since by Assumption~\ref{assum:2} the matrix $A^1$ is irreducible, $\beta_1^1>0$, and from statement~\ref{r2} we know that $\bar{x}^1 \gg \textbf{0}$, it must be that $(Q_2\bar{x}^1)_i<0$. Suppose secondly  that $B_i^1 \neq \mathbf{0}$. Since from statement~\ref{r2} we know that $\bar{x}^1_i \geq \frac{1}{2}$, it follows that $1-2\bar{x}^1_i <0$; thus implying that $(Q_2\bar{x}^1)_i<0$. Note that the choice of index $i$ was arbitrary, and therefore again, we have  $(Q_2\bar{x}^1)_i<0$ for all $i \in [n]$. Hence, since $\bar{J}_{11}\bar{x}^1 =Q_2\bar{x}^1$, it follows that  $(\bar{J}_{11}\bar{x}^1)_i<0$ for all $i \in [n]$. Note that Assumptions~\ref{assum:1} and~\ref{assum:2} guarantee that the matrices $Q_1$ and $Q_2$ are irreducible Metzler matrices; hence, the matrix $\bar{J}_{11}$ is an irreducible Metzler matrix. Therefore, from Lemma~\ref{lem:ratzner}, it must be that the matrix $\bar{J}_{11}$ is Hurwitz.\\

\noindent Turning our attention to the matrix $\bar{J}_{22}$,
 consider the matrices $\beta_1^2(D^2)^{-1}A^2$ and  $\beta_1^2(D^2)^{-1}(I-\bar{X}^1)A^2$. From Assumption~\ref{assum:1}, 
it is clear that $\beta_1^2(D^2)^{-1}A^2$ is a 
 nonnegative matrix. Since, from statement~\ref{r2}, $\bar{x}^1$ satisfies $\textbf{0} \ll \bar{x}^1 \ll \textbf{1}$, it is also clear that  $\beta_1^2(D^2)^{-1}(I-\bar{X}^1)A^2$ is 
a  nonnegative matrix. Furthermore, we also immediately obtain the following:
$$\beta_1^2(D^2)^{-1}(I-\bar{X}^1)A^2 < \beta_1^2(D^2)^{-1}A^2.$$ Therefore, since the spectral radius of a nonnegative matrix decreases monotonically with a decrease in any entry of said matrix, 
 it follows that 
$\rho(\beta_1^2(D^2)^{-1}(I-\bar{X}^1)A^2) \leq \rho(\beta_1^2(D^2)^{-1}A^2)$. By assumption, $\rho(\beta_1^2(D^2)^{-1}A^2)<1$, which implies that $\rho(\beta_1^2(D^2)^{-1}(I-\bar{X}^1)A^2)<1$, and consequently, from Lemma~\ref{lem:eigspec}, we have that $s(-D^2+(I-\bar{X}^1)A^2)<0$. Therefore, since $J(\bar{x}^1, \textbf{0})$ is block upper triangular, and since we have already established that $\bar{J}_{11}$ is Hurwitz, it follows that $s(J(\bar{x}^1, \textbf{0}))<0$. Local 
exponential 
stability of $(\bar{x}^1, \textbf{0})$, then, follows from \cite[Theorem 4.15 and Corollary~4.3]{khalil2002nonlinear}.\\

\noindent \textit{Proof of statement~\ref{r4}:} The proof is analogous to that of statement~\ref{r3}.~\qed\\

Proposition~\ref{prop:tristable} answers question~\ref{qs1} raised in Section~\ref{sec:prob:formulation}. Proposition~\ref{prop:tristable} guarantees the existence and simultaneous local exponential stability of three equilibria, whereas \cite[Theorem~5.3]{cui2023general}, assuming that an endemic equilibrium exists, guarantees its local stability. Furthermore, the possibility of the DFE being locally stable simultaneously is alluded to; see \cite[Remark~10]{cui2023general}. 
 On the other hand, for a parameter regime different from the one covered in Proposition~\ref{prop:tristable}, assuming that the terms corresponding to HOI are sufficiently small, \cite[Theorem~5.3]{cui2023general} secures global stability of the endemic equilibrium.
 The following remarks are in order.

\begin{remark}\label{rem:tristable}
Proposition~\ref{prop:tristable} sheds light on an interesting phenomenon that bivirus spread over hypergraph exhibits (but bivirus spread over a normal graph does not exhibit): identification of a parameter regime that permits three equilibria, namely the DFE and the two boundary equilibria, to be simultaneously stable. This is an extension to the single virus case studied in \cite{cisneros2021multigroup}, which permitted the simultaneous stability of the DFE and (since there is only one in this case) an endemic equilibrium.
\end{remark}

\begin{remark}\label{rem:necessity of spectral radiuscriteria}
It is known that, assuming $\beta_2^k=0$ for $k=1,2$, the condition $\rho(\beta_1^k(D^k)^{-1}A^k)\leq 1$  guarantees that the DFE is the only equilibrium of system~\eqref{eq:bivirus:hypergraph}; see \cite[Lemma~2]{pare2021multi}. However, as Proposition~\ref{prop:tristable} shows, that is not necessarily true when considering bivirus SIS spread over hypergraphs.
\end{remark}

\noindent Proposition~\ref{prop:tristable} guarantees existence of boundary equilibria for the case when \break  $\rho(\beta_1^k(D^k)^{-1}A^k)<1$. It is natural to ask if one is assured of existence even if the spectral radii of relevant quantities are larger than one. The following proposition addresses this issue.
\begin{prop}\label{prop:largerthanone}
Consider system~\eqref{eq:bivirus:hypergraph} under Assumptions~\ref{assum:1} and~\ref{assum:2}. Suppose that, for all $k \in [2]$, $\rho(\beta_1^k(D^k)^{-1}A^k)>1$. Then system~\eqref{eq:bivirus:hypergraph} has at least three equilibria, namely the DFE, 
a single virus  endemic equilibrium corresponding to virus~$1$ $(\bar{x}^1, \textbf{0})$, and 
a single virus  endemic equilibrium corresponding to virus~$2$ $(\textbf{0}, \bar{x}^2)$. Furthermore, if $s(-D^i +\beta_1^i(I-\bar{X}^k)A^i)>0$ for $i,k \in [2]$ such that $i\neq k$, then the 
 equilibrium points $(\bar{x}^1, \textbf{0})$ and $(\textbf{0},\bar{x}^2)$ 
are unstable. 
 \end{prop}
\noindent \textit{Proof:} Observe that the DFE is always an equilibrium of system~\eqref{eq:bivirus:hypergraph}. Suppose that for some $k \in [2]$, $\rho(\beta_1^k(D^k)^{-1}A^k)>1$, then from \cite[Theorem~5.1, statement~vii)]{cisneros2021multigroup} we know that there exists an endemic equilibrium, $\bar{x}^k$, where $\textbf{0} \ll \bar{x}^k \ll \mathbf{1}$. Since, by assumption, $\rho(\beta_1^k(D^k)^{-1}A^k)>1$ for all $k \in [2]$, it is also immediate that there exist equilibria, $(\bar{x}^1, \textbf{0})$ and $(\textbf{0}, \bar{x}^2)$, where $\textbf{0} \ll \bar{x}^k \ll \mathbf{1}$, for $k=1,2$. \\
It can be verified that $J(\bar{x}^1, \textbf{0})$ is block upper triangular, with the matrix $-D^1 +\beta_1^1(I-\bar{X}^2)A^1$ being one of the blocks along the diagonal. By assumption, $s(-D^1 +\beta_1^1(I-\bar{X}^2)A^1)>0$, which implies that $s(J(\bar{x}^1, \textbf{0}))>0$. Consequently, instability of $(\bar{x}^1, \textbf{0})$ follows from \cite[Theorem~4.7, statement ~ii)]{khalil2002nonlinear}. The instability of $(\textbf{0}, \bar{x}^2)$ can be shown analogously, thus completing the proof.~\qed
 \begin{remark}\label{rem:computation:endemic:equilibria}
Proposition~\ref{prop:largerthanone} (resp. Proposition~\ref{prop:tristable})  guarantees the existence (resp. existence and local exponential stability) of the equilibrium points, $(\bar{x}^1, \textbf{0})$ and $(\textbf{0}, \bar{x}^2)$. It turns out that it is possible to compute these points iteratively; see \cite[Theorem~5.3]{cisneros2021multigroup}.
\end{remark}

\section{(Non)existence of Coexistence equilibria}\label{sec:coexistence}
This section identifies sufficient conditions for the existence (resp. nonexistence) of coexistence equilibria. Specifically, for investigating existence, we consider two parameter regimes, viz. for $k=1,2$, i) $s(-D^k +\beta_i^k A^k)>0$, and ii) $s(-D^k +\beta_i^k A^k)<0$. Further, for the parameter regime i), we consider two stability configurations of the boundary equilibria, viz. a) both being unstable and b) both being stable; for parameter regime ii), we consider the case where both boundary equilibria are stable.  
\begin{prop}\label{prop:1:suff:coexistence}
    Consider system~\eqref{eq:bivirus:hypergraph} under Assumptions~\ref{assum:1} and~\ref{assum:2}. Let $(\bar{x}^1, \textbf{0})$ and $(\textbf{0}, \bar{x}^2)$ denote a single-virus endemic equilibrium corresponding to virus~1 and virus~2, respectively.
 Suppose that the following conditions are satisfied:
\begin{enumerate}[label=\roman*)]
\item\label{a11} $s(-D^1 + \beta_1^1A^1)>0$;
\item\label{a21} $s(-D^2 + \beta_1^2A^2)>0$;
\item\label{a31} $s(-D^1 +\beta_1^1(I-\bar{X}^2)A^1)>0$; and
\item\label{a41} $s(-D^2 +\beta_1^2(I-\bar{X}^1)A^2)>0$.
\end{enumerate}
Then there exists at least one equilibrium of the form $(\hat{x}^1, \hat{x}^2)$ such that $\textbf{0} \ll \hat{x}^1, \hat{x}^2 \ll \textbf{1}$ and $\hat{x}^1+ \hat{x}^2 \ll \textbf{1}$.
\end{prop}

Before proving the claim in Proposition~\ref{prop:1:suff:coexistence}, we need the following background material.
In line with the terminology of \cite{hofbauer1990index}, given an equilibrium point of system~\eqref{eq:bivirus:hypergraph}, we classify the same
as saturated or unsaturated. 
We say that an equilibrium is 
saturated  (resp. strictly saturated) if the diagonal block corresponding to the zero entries of said equilibrium  possibly has a single eigenvalue at the origin (resp. has every eigenvalue to be strictly less than zero)
and unsaturated otherwise~\cite{hofbauer1990index}. 
A boundary equilibrium of~\eqref{eq:bivirus:hypergraph} is saturated if and only if said boundary equilibrium is locally exponentially stable; this follows immediately by noting the structure of the Jacobian matrix, evaluated at a boundary equilibrium, see \eqref{jac:boundary}.
The definition also implies that every fixed point in the interior of $\mathcal D$, irrespective of its stability properties, is saturated \cite{hofbauer1990index}; therefore,  from Lemma~\ref{lem:equi_non-zero_nonone:1}, we have that every coexistence equilibrium of system~\eqref{eq:bivirus:hypergraph} is saturated.

\noindent \textit{Proof:} Assumptions~\ref{a11} and~\ref{a21} of Proposition~\ref{prop:1:suff:coexistence} guarantee existence of boundary equilibria, $(\bar{x}^1, \textbf{0})$ and $(\textbf{0}, \bar{x}^2)$; see Proposition~\ref{prop:largerthanone}. Observe that \cite[Lemma~5.1]{cui2023general} guarantees that, for each $k \in [2]$, $x^k(0)\geq\textbf{0}$ implies that $x^k(t)\geq\textbf{0}$ for all $t \in \mathbb{R}_{\geq 0}$, and that the set $\mathcal D$ (which is compact) is forward invariant. Therefore, from \cite[Theorem~2]{hofbauer1990index}, it follows that system~\eqref{eq:bivirus:hypergraph} has at least one saturated fixed point. There are two cases to consider.\\
Case 1: Suppose the aforementioned saturated fixed point is in the interior of $\mathcal D$. 
 Note that any fixed point in the interior of $\mathcal D$ is 
of the form $(\hat{x}^1, \hat{x}^2)$, where $\textbf{0} \ll (\hat{x}^1, \hat{x}^2) \ll \textbf{1}$, thus implying that $(\hat{x}^1, \hat{x}^2)$ is a coexistence equilibrium. From Lemma~\ref{lem:equi_non-zero_nonone:1}, it must necessarily satisfy $\textbf{0} \ll (\hat{x}^1, \hat{x}^2) \ll \textbf{1}$, and $\hat{x}^1+ \hat{x}^2 \ll \textbf{1}$. \\
Case 2: Suppose, but we will demonstrate a contradiction, that there are no fixed points in the interior of $\mathcal D$. This implies that there must be a saturated fixed point on the boundary of $\mathcal D$  \cite{hofbauer1990index}. Therefore,   at least one of the single-virus boundary equilibria is saturated.\\
However, from Proposition~\ref{prop:largerthanone}, it is clear that assumptions ~\ref{a31} and~\ref{a41} 
guarantee that the boundary equilibria are unstable; thus implying that they are unsaturated, and the contradiction is obtained. ~\qed

Proposition~\ref{prop:1:suff:coexistence} is 
implied by \cite[Theorem~5.4]{cui2023general}, which, assuming $\beta_2^k=0$ for $k=1,2$, is the same as 
\cite[Theorem~5]{axel2020TAC}. 
The proof technique in \cite[Theorem~5.4]{cui2023general} is quite involved since it primarily relies on fixed point mapping, Perron Frobenius theory, etc. Our proof is significantly shorter. Note that \cite[Theorem~2]{hofbauer1990index} is a key ingredient of our proof strategy. In light of Theorem~\ref{thm:finiteness}, one could perhaps leverage \cite[Theorem~2]{hofbauer1990index}  to obtain a lower bound on the number of coexistence equilibria for the stability configuration of boundary equilibria given in Proposition~\ref{prop:1:suff:coexistence}, as has been done for classic bivirus networked SIS models; see \cite[Corollary~3.9, statement 2]{anderson2023equilibria}. Subsequently, one could possibly exploit the properties of MDS to conclude that there must exist a locally exponentially stable coexistence equilibrium. 

Observe that in Propostion~\ref{prop:1:suff:coexistence} the demonstration of
 the existence of a coexistence equilibrium point $(\hat{x}^1, \hat{x}^2)$ relies on the assumption that both boundary equilibria are unstable. We now present a different condition that also guarantees the existence of a coexistence equilibrium point $(\hat{x}^1, \hat{x}^2)$ even when both boundary equilibria are stable. 

\begin{theorem}\label{prop:suff:coexistence:both stable}
 Consider system~\eqref{eq:bivirus:hypergraph} 
under Assumptions~\ref{assum:1} and~\ref{assum:2}.  Let $(\bar{x}^1, \textbf{0})$ and $(\textbf{0}, \bar{x}^2)$ denote a single-virus endemic equilibrium corresponding to virus~1 and virus~2, respectively. Suppose that the following conditions are satisfied:
\begin{enumerate}[label=\roman*)]
\item\label{a1} $s(-D^1 + \beta_1^1A^1)>0$;
\item\label{a2} $s(-D^2 + \beta_1^2A^2)>0$;
\end{enumerate}
Suppose that both $(\bar{x}^1, \textbf{0})$ and $(\textbf{0}, \bar{x}^2)$ are locally exponentially stable.
Then there exists at least one 
equilibrium of the form $(\hat{x}^1, \hat{x}^2)$ such that $\textbf{0} \ll \hat{x}^1, \hat{x}^2 \ll \textbf{1}$ and $\hat{x}^1+ \hat{x}^2 \ll \textbf{1}$, such that $(\hat{x}^1, \hat{x}^2)$ is either neutrally stable or unstable.\footnote{Assuming that equilibria of system~\eqref{eq:bivirus:hypergraph} are hyperbolic, a stronger conclusion can be drawn: for generic parameter matrices, the coexistence equilibrium is unstable.} 
\end{theorem}
\noindent \textit{Proof:} By assumption,  $s(-D^k + \beta_1^kA^k)>0$ for $k=1,2$. Therefore, from Proposition~\ref{prop:largerthanone}, it follows that there exists a single-virus endemic equilibrium corresponding to virus~1, $\bar{x}^1 \gg \textbf{0}$, and a single-virus endemic equilibrium corresponding to virus~2, $\bar{x}^2 \gg \textbf{0}$. By assumption, both $(\bar{x}^1, \textbf{0})$ and $(\textbf{0}, \bar{x}^2)$ are locally exponentially stable.\\
The condition $s(-D^1 + \beta_1^1A^1)>0$ implies that the origin is unstable; this can be observed from the proof of statement~i) in Proposition~\ref{prop:tristable}. We are left to show that the stable manifold of the origin does not lie in the interior of $\mathcal D$. We will rely on the proof technique of \cite[Lemma~3.8]{anderson2023equilibria}. It suffices to show that for the (linear) system 
\begin{equation}
\begin{bmatrix}
\dot{x}^1\\ \dot{x}^2
\end{bmatrix} =
\begin{bmatrix}
-D^1+\beta_1^1A^1 & \textbf{0} \\
\textbf{0} & -D^2+\beta_1^2A^2
\end{bmatrix}
\begin{bmatrix}
x^1\\ x^2
\end{bmatrix}\label{eq:linear}
\end{equation}
no trajectory starting in the interior of $\mathcal D$ converges to the origin. First, consider $x^1(t)$. Let $w^\top$ be the left eigenvector associated with $s(-D^1+\beta_1^1A^1)$ so all its entries sum to one. Define $z:=w^\top x^1$, and observe that $\dot{z}=w^\top \dot{x}^1$, which, from~\eqref{eq:linear}, further implies that $\dot{z}=s(-D^1+\beta_1^1A^1)z$. Since, by assumption, $s(-D^1+\beta_1^1A^1)>0$, it is clear that the projection onto $w$ (which is a positive vector) of the points of~\eqref{eq:linear} in the interior of $\mathcal D$ is away from $x^1=\textbf{0}$. An analogous argument can be made for $x^2(t)$, since, by assumption, $s(-D^2+\beta_1^2A^2)>0$. Therefore, the stable manifold of the origin does not lie in the interior of $\mathcal D$. Consequently, since we know that system~\eqref{eq:bivirus:hypergraph} is monotone (see \cite[Theorem~5.5]{cui2023general}) and the monotone condition $\bar x^1\gg{\bf{0}}, {\bf{0}}\ll \bar x^2$ relates the two exponentially stable equilibrium points, from \cite[Proposition~2.9]{smith1988systems} it follows that there exists an equilibrium point of the form $(\hat{x}^1, \hat{x}^2)$ such that $\textbf{0} \ll \hat{x}^1, \hat{x}^2 \ll \textbf{1}$ and $\hat{x}^1+ \hat{x}^2 \ll \textbf{1}$. Furthermore, the point $(\hat{x}^1, \hat{x}^2)$ satisfies $s(J(\hat{x}^1, \hat{x}^2))\geq 0$, thus concluding the proof.~\qed\\
 

Proposition~\ref{prop:1:suff:coexistence} and Theorem~\ref{prop:suff:coexistence:both stable} partially answer question~\ref{qs2} raised in Section~\ref{sec:prob:statement}. Observe that neither of these results covers the case where one boundary equilibrium is locally exponentially stable, and the other is unstable.\\
We next consider a different parameter regime, namely $s(-D^k+\beta_1^k A^k)<0$, and identify a sufficient condition for the existence of an unstable coexistence equilibrium. We have the following result.


\begin{prop}\label{prop:2:suff:coexistence:less than 1}
Consider system~\eqref{eq:bivirus:hypergraph} 
under Assumptions~\ref{assum:1} and~\ref{assum:2}. Define, for $k=1,2$, $\mathbf{1}_{B^k} \in \{0,1\}^n$ by $(\mathbf{1}_{B^k})_i=1$ if $B_i^k \neq \mathbf{0}$; otherwise $(\mathbf{1}_{B^k})_i=0$. Suppose that the following conditions are fulfilled for $k=1,2$ :
    \begin{enumerate}[label=\alph*)]
    \item\label{q11} $\rho(\beta_1^k(D^k)^{-1}A^k)<1$, and
    \item\label{q22} $\min\limits_{i. s.t. B_i^k \neq \mathbf{0}}\big(\frac{\beta_1^k}{\delta_i^k} (A^k\mathbf{1}_{B^k})_i + \frac{\beta_2^k}{2\delta_i^k}\mathbf{1}_{B^k}^\top B_i \mathbf{1}_{B^k} \big)>2$.
    \end{enumerate}
Then there exists at least one  equilibrium of the form $(\hat{x}^1, \hat{x}^2)$ such that $\textbf{0} \ll \hat{x}^1, \hat{x}^2 \ll \textbf{1}$ and $\hat{x}^1+ \hat{x}^2 \ll \textbf{1}$ that is either neutrally stable or unstable.
\end{prop}
\noindent \textit{Proof:} Suppose that the conditions in Proposition~\ref{prop:2:suff:coexistence:less than 1} are fulfilled.  Therefore, it follows that there exist boundary equilibria $(\bar{x}^1, \textbf{0})$ and $(\textbf{0}, \bar{x}^2)$, and that both  are locally exponentially stable; see statements~\ref{r3} and~\ref{r4} in Proposition~\ref{prop:tristable}. Therefore, since we know that system~\eqref{eq:bivirus:hypergraph} is monotone (see \cite[Theorem~5.5]{cui2023general}), from \cite[Proposition~2.9]{smith1988systems} it follows that there exists (at least) one equilibrium point of the form $(\hat{x}^1, \hat{x}^2)$ such that $\textbf{0} \ll \hat{x}^1, \hat{x}^2 \ll \textbf{1}$ and $\hat{x}^1+ \hat{x}^2 \ll \textbf{1}$. Furthermore, $s(J(\hat{x}^1, \hat{x}^2))\geq 0$, thus delivering the claim.~\qed

Note that Proposition~\ref{prop:2:suff:coexistence:less than 1} guarantees the existence of at least one coexistence equilibrium. Given that system~\eqref{eq:bivirus:hypergraph} is monotone, and since, from Theorem~\ref{thm:finiteness}, it is known that for each of the equilibrium points the associated Jacobian is nonsingular, the conditions in Proposition~\ref{prop:2:suff:coexistence:less than 1} 
guarantee the existence of an odd number of coexistence equilibria, each of which must be unstable. The proof for the same follows from a Brouwer degree argument; see \cite{smith1986competing}. In fact, for the special case where $\beta_2^k=0$ for $k=1,2$, for the same stability configuration as in Theorem~\ref{prop:suff:coexistence:both stable} and
Proposition~\ref{prop:2:suff:coexistence:less than 1}, a lower bound on the number of coexistence equilibria has been recently provided; see \cite[Corollary~3.9, statement 3]{anderson2023equilibria}.

\section{Numerical Examples}\label{sec:simulation}

{\color{black}We present a series of simulations highlighting interesting phenomena that can emerge when HOIs are incorporated. We use the following bivirus system with HOIs. The network has $n=5$ nodes, and we set $D^1 = D^2 = I$. The pairwise interactions are captured by two-cycle graphs with self-loops, with infection matrices:
\begin{equation}\label{eq:A_simulation}
	A^1 = \begin{bmatrix}
		1 & 0 & 0 & 0 &1 \\ 1 &  1 & 0 & 0 & 0 \\ 0 & 1 & 1 & 0 & 0 \\ 0 & 0 & 1 & 1 & 0 \\ 0 & 0 & 0 & 1 & 1
	\end{bmatrix} \,, \qquad A^2 = (A^1)^\top.
\end{equation}
\normalsize 
The 
HOI are captured by the following set of hyperedges with unit weight:
\begin{align*}
	\text{virus } 1:  (1,2,3), (2,3,1), (3,2,1), (1,4,5), (4,5,1), (5,4,1) \\
	\text{virus } 2:  (1,2,4), (2,4,1), (4,2,1), (1,3,5), (3,5,1), (5,3,1).
\end{align*}
In other words, this corresponds to the following $b^k_{ij\ell}$ entries being equal to $1$, with all other entries of $B^k_{i}$ equal to $0$: $b^1_{123}$, $b^1_{231}$, $b^1_{321}$, $b^1_{145}$, $b^1_{451}$, $b^1_{541}$, and $b^2_{124}$, $b^2_{241}$, $b^2_{421}$, $b^2_{135}$, $b^2_{351}$, $b^2_{531}$. In our simulations, we randomly sample $x^k_i(0)$ from a uniform distribution $(0,1)$, and then normalize the vectors $x^1(0)$ and $x^2(0)$ to ensure that $(x^1(0),x^2(0)) \in \text{int}(\mathcal{D})$. The $\beta^k_i$ are varied to yield different stability properties for the system in \eqref{eq:bivirus:hypergraph}.

\textit{Example 1:} We set $\beta^1_1 = \beta^1_2 = 0.2$ and $\beta^2_1 = \beta^2_2 = 5$. This ensures the inequalities of both conditions for Proposition~1 are satisfied. As can be observed from Fig.~\ref{fig:prop1_dfe}, for initial conditions close to the DFE, the trajectories converge to the locally exponentially stable DFE, $(x^1 = {\bf 0}, x^2 = {\bf 0})$. In Figs.~\ref{fig:prop1_v1} and \ref{fig:prop1_v2}, the initial conditions are further in the interior of $\mathcal{D}$, and depending on the particular initial condition, we observe convergence to a boundary equilibrium where one of the two viruses is extinct, $(\bar x^1, {\bf 0})$ or $({\bf 0}, \bar x^2)$ for some positive $\bar x^1 > 0.5 \times {\bf 1}$ and $\bar x^2 > 0.5 \times {\bf 1}$. That is, both boundary equilibria are simultaneously locally exponentially stable. Interestingly, without HOIs, it is impossible for a bivirus system to have the DFE, $(\bar x^1, {\bf 0})$, and $({\bf 0}, \bar x^2)$ all locally exponentially stable~\cite[Section~E]{ben:lcss}.

\begin{figure*}
\begin{minipage}{0.25\linewidth}
\centering
\subfloat[]{\includegraphics[width=\columnwidth]{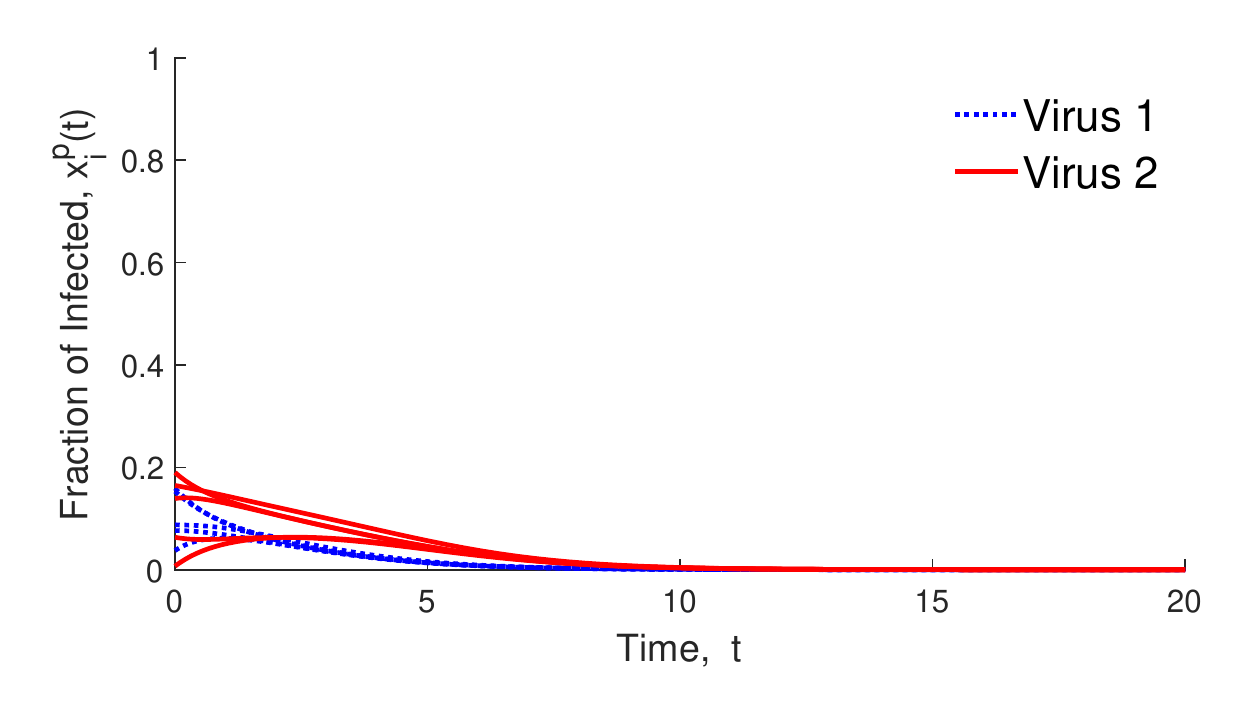}\label{fig:prop1_dfe}}
\end{minipage}
\hfill
\begin{minipage}{0.25\linewidth}
\centering\subfloat[]{\includegraphics[width=\columnwidth]{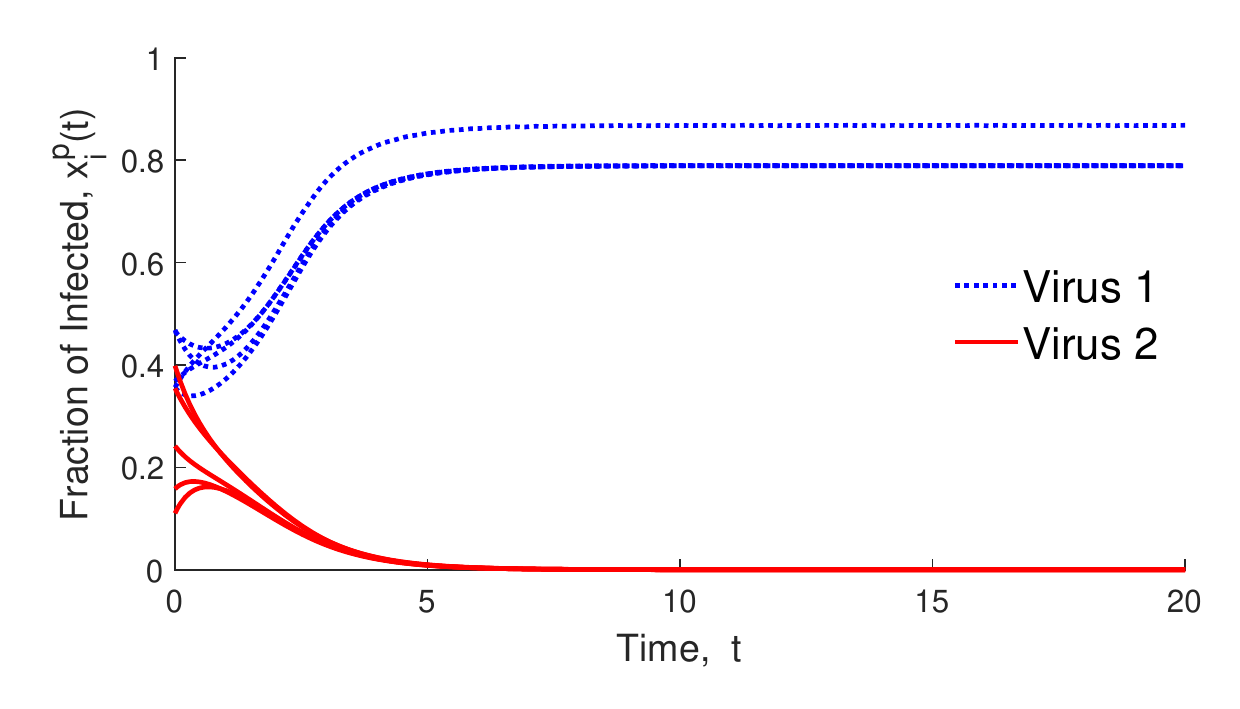}\label{fig:prop1_v1}}
\end{minipage}
\hfill
\begin{minipage}{0.25\linewidth}
\centering
\subfloat[]{\includegraphics[width=\columnwidth]{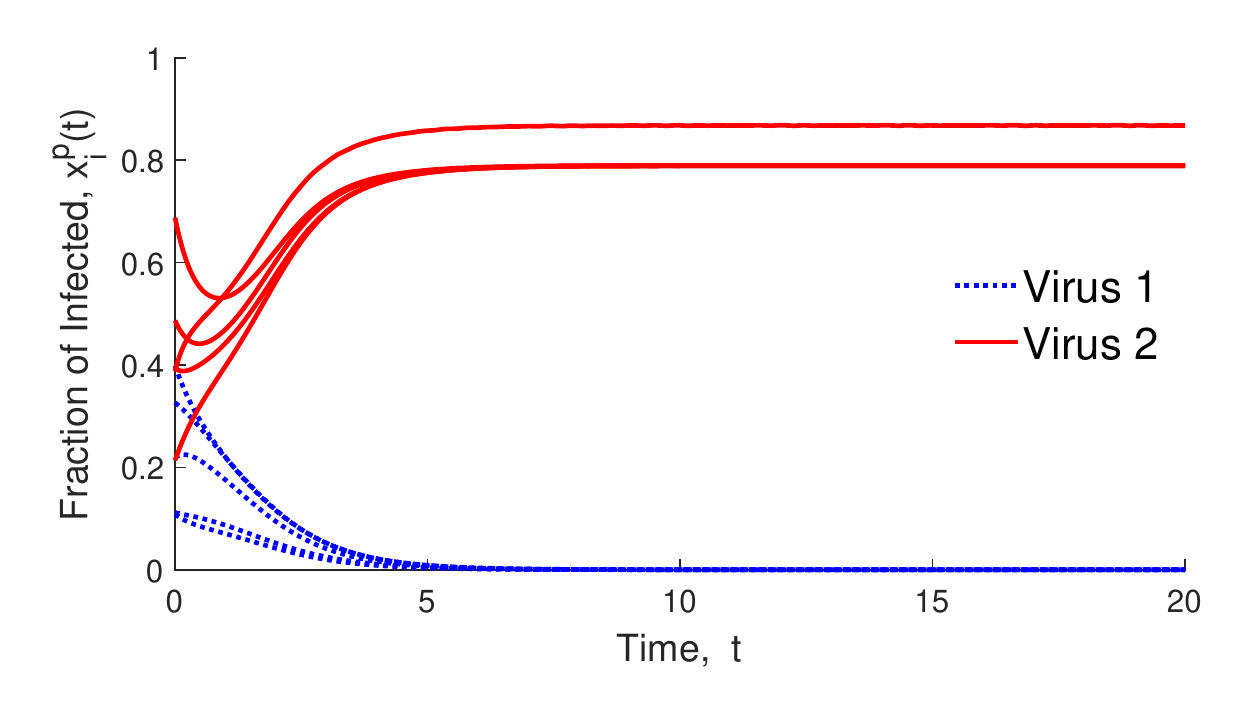}\label{fig:prop1_v2}}
\end{minipage}
\caption{Trajectories of the system \eqref{eq:bivirus:hypergraph}, for different initial conditions in \textit{Example 1}. }\label{fig:simulations_1}
\end{figure*}

\textit{Example 2:} We set $\beta^1_1 = \beta^1_2 = 2$ and $\beta^2_1 = 3$ and $\beta^2_2 = 2.4$. As illustrated in Figs.~\ref{fig:rho_doublestable_ex1} and \ref{fig:rho_doublestable_ex2}, there are two locally exponentially stable two boundary equilibria $(\bar x^1, {\bf 0})$ or $({\bf 0}, \bar x^2)$, and we converge to either depending on the initial conditions. However, the DFE is unstable, and no trajectories in $\mathcal{D}$ converge there except if one starts at the DFE. This simulation highlights are interesting observation: for a standard bivirus system with no HOIs, examples of systems with two locally stable boundary equilibria have not been identified until recently and are not straightforward to construct \cite{anderson2023equilibria,ye2021_bivirus_outcomes}. 

\begin{figure}
\begin{minipage}{0.49\linewidth}
\centering
\subfloat[Example~2]{\includegraphics[width=\columnwidth]{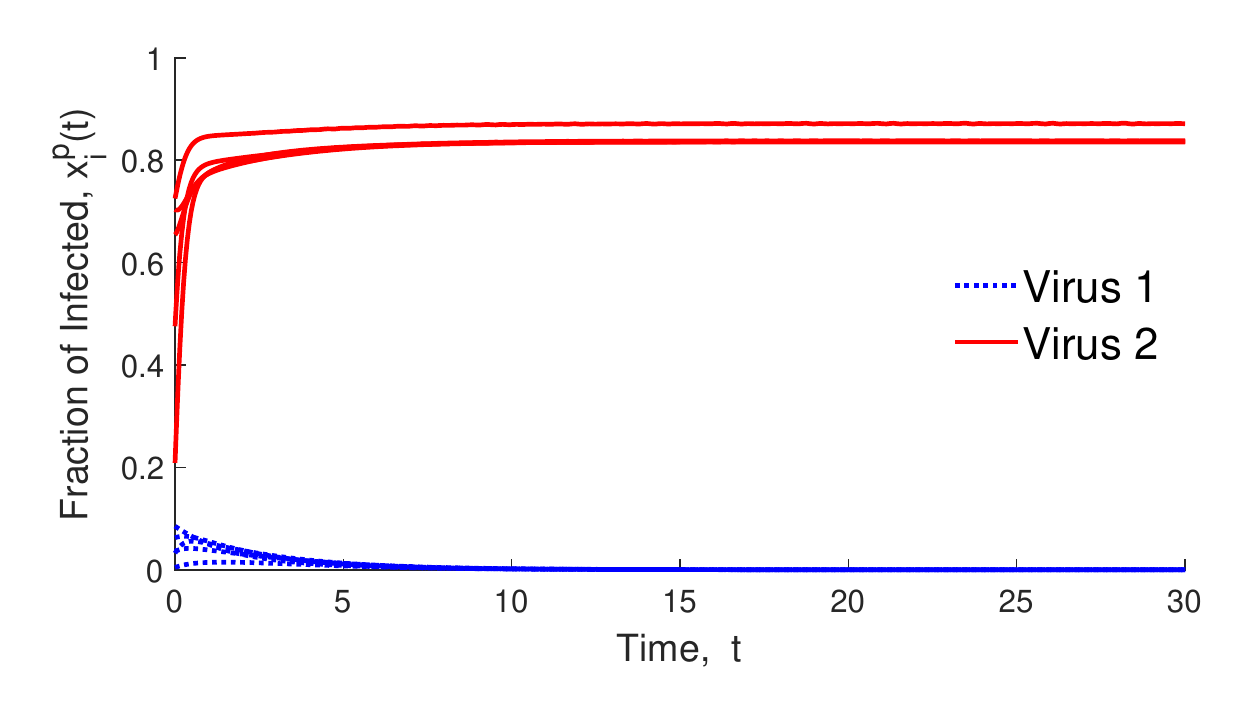}\label{fig:rho_doublestable_ex1}}
\end{minipage}
\hfill
\begin{minipage}{0.49\linewidth}
\centering\subfloat[Example~2]{\includegraphics[width=\columnwidth]{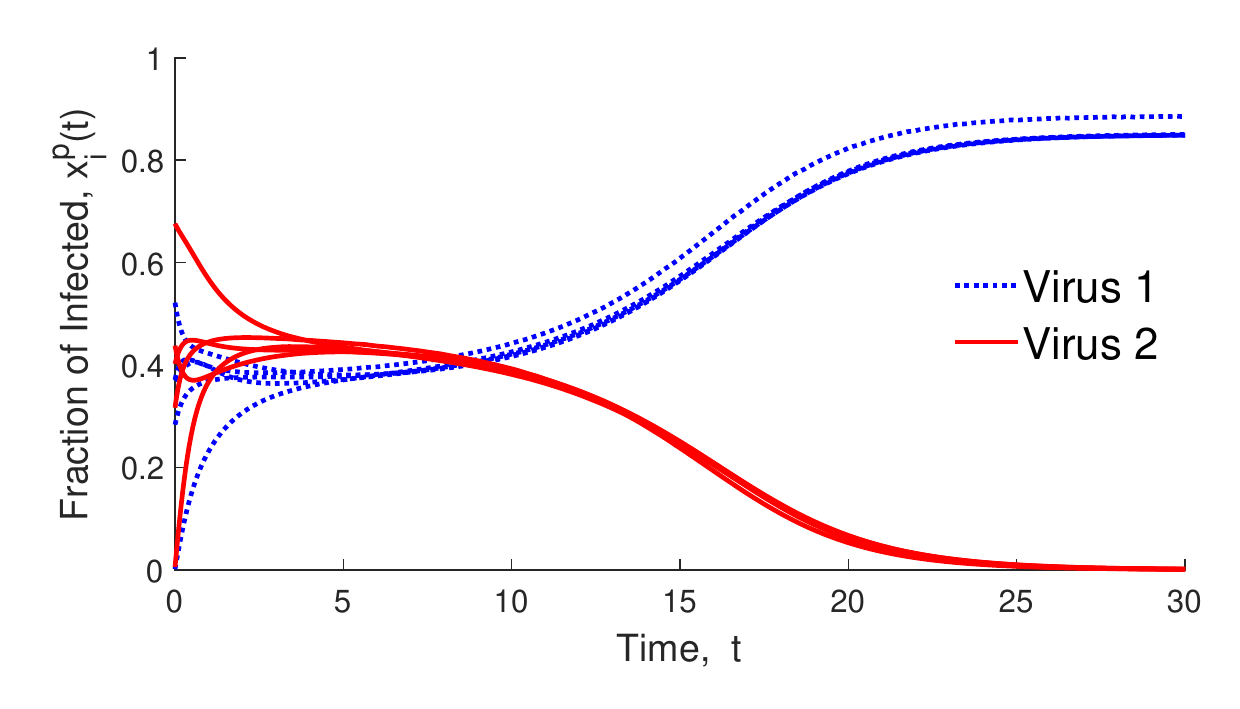}\label{fig:rho_doublestable_ex2}}
\end{minipage}
\vfill
\caption{Trajectories of the system \eqref{eq:bivirus:hypergraph}, for different simulation parameters and initial conditions. \ref{fig:rho_doublestable_ex1} and \ref{fig:rho_doublestable_ex2} correspond to different initial conditions for \textit{Example 2}. 
}\label{fig:simulations_2}
\end{figure}


We conclude by remarking that, for each of the simulations presented, the bivirus system exhibits dynamical phenomena when HOIs are present that is not observed when HOIs are not present. In other words, HOIs unlock new possibilities in the competition of two viruses spreading over networks and suggest a significant amount of understanding remains to be unveiled.
}

\section{Conclusion}\label{sec:conclusion}
This paper analyzed a networked competitive bivirus SIS model that also accounts for the possibility of HOI among the nodes. By taking recourse to the Parametric Transversality Theorem of differential topology, we showed that the bivirus system with HOI has, for generic parameter values, a finite number of equilibria. Furthermore,  the Jacobian matrices associated with each of the equilibria are nonsingular. This finding, coupled with the knowledge that the system is monotone, enabled us to establish that the typical behavior that our system exhibits is convergence to some equilibrium. Subsequently, we identified a parameter regime that ensures the existence of multiple boundary equilibria and simultaneous stability of the same along with that of the DFE. For the special case where only one virus is circulating in the meta-population, we guarantee the existence and local stability of an endemic equilibrium; our result does not impose any restrictions on the model parameters besides those covered by Assumptions~1 and~2. Thereafter, for different parameter regimes, we identified conditions that guarantee the existence of a coexistence equilibrium. 


\bibliography{ReferencesRice}

\section*{Appendix}
\subsection*{Proof of Theorem~\ref{thm:finiteness}}
 We treat first the case where the $A^k$ and $B^k_i$ are fixed. Consider \eqref{eq:bivirus:hypergraph} in the shorthand form $\dot x=f(x)$, and to emphasize the possibility of variations in the entries of $D^k$, show this explicitly by writing the equations as 
\begin{equation}
    \dot x=f_{\delta}(x,\delta)
\end{equation}
where $\delta$ is a vector containing all the diagonal entries of the $D^k$. It is easily established that 
\begin{equation}
    \frac{\partial f_{\delta}(x^1,x^2,D^1,D^2)}{\partial \delta}=\frac{\partial f_{\delta}(x^1,x^2,D^1,D^2)}{\partial(D^1,D^2)}
 =\begin{bmatrix}
        X^1&0\\0&X^2
    \end{bmatrix} \nonumber
\end{equation} 
At a coexistence equilibrium, 
this matrix has full row rank. Consequently, the matrix $\frac{\partial f_{\delta}(x,\delta)}{\partial (x,\delta)}$ (which is obtained simply by adding further columns) also has full row rank at a coexisitence equilibrium.

Similarly, in preparation for considering the second claim of the theorem, we can consider $f$ depending explicitly on the possibly variable $A^k$. There are a finite number of patterns of zero entries for the $A^k$ for which the $A^k$ are irreducible. For each matrix separately, arbitrarily choose any one such pattern, and let $\alpha^k$ denote the vector of nonzero entries of the $A^k$; also set $\alpha=[(\alpha^1)^{\top}\;(\alpha^2)^{\top}]^{\top}$. Denote the associated $f$ as $f_{\alpha}(x,\alpha)$. Then one can show that
\footnotesize 
\begin{equation}
    \frac{\partial f_{\alpha}(x,\alpha)}{\partial\alpha}=\begin{bmatrix}
        I-X^1-X^2&0\\0&I-X^1-X^2\end{bmatrix}
        \begin{bmatrix}
            \frac{\partial(A^1x^1)}{\partial \alpha^1}&0\\0&\frac{\partial (A^2x^2)}{\partial \alpha^2}
        \end{bmatrix}\label{eq:varying:Ak}
\end{equation}
\normalsize 
A detailed calculation as set out in \cite{anderson2023equilibria} shows that the second matrix in the product 
has full row rank. Examination of the system equation \eqref{eq:bivirus:hypergraph} immediately reveals that no equilibrium can occur where any diagonal entry of $I-X^1-X^2=0$, which means that the first matrix of the product is nonsingular. Hence the matrix on the left has full row rank at any equilibrium. The matrix $\frac{\partial f_{\alpha}(x,\alpha)}{\partial (x,\alpha)}$, obtained by the addition of further columns to $\frac{\partial f_{\alpha}(x,\alpha)}{\partial\alpha}$, then has full row rank. 

We now complete the proof for the first part of the theorem. (Proof of the second part is virtually identical). Without loss of generality, suppose that for some $\bar {\delta}$, all diagonal entries $\delta_i$ of the $D^k$ satisfy $0<\bar {\delta}<\delta_i<\bar {\delta }^{-1}$. Denote the set of such $\delta$ as  $\mathcal D$. Call the set of $x$ in the interior of the region of interest $\mathcal X$. Then $\mathcal X \times \mathcal D$ is a manifold, being a product of open sets, and $f_\delta(x,\delta)$ is a mapping of the manifold $\mathcal X\times\mathcal D$. Let the image be denoted by $\mathcal Y$, and let $\mathcal Z$ be the submanifold of $\mathcal Y$ consisting of the single point ${\bf{0_{2n}}}$. (The set $f_{\delta}^{-1}(\mathcal Z)$ is the set of zeros of $f$.)  Since the Jacobian of $f_{\delta}$ with respect to $(x,\delta)$ has full row rank on this set, this means that the map $f_{\delta}$ is transversal to $\mathcal Z$. By the Parametric Transversality Theorem, this means that for all choices of $\delta=\delta^*$ excluding a set of measure zero, the Jacobian $\frac{\partial f_{\delta^*}(x,\delta^*)}{\partial x}$ will be nonsingular at a preimage of $\mathcal Z$, i.e at a zero. This is equivalent to saying that the zeros are nondegenerate. The boundedness of the region $\mathcal X$ then guarantees that there can only be a finite number of such zeros. 
\end{document}